\documentclass[preprint,12pt]{elsarticle}

\usepackage{amsmath,amssymb,amsfonts}
\usepackage{graphicx}
\usepackage{textcomp}
\usepackage{booktabs}
\usepackage{threeparttable}
\usepackage{multirow}
\usepackage{tabularx}
\usepackage{algorithm, algorithmic}
\usepackage{comment}

\begin{document}

\begin{frontmatter}



\title{Optimizing Surgical Plans for Parenchyma-Sparing Liver Resections through Contour-Guided Resection and Surface Approximation} 

\author[1,2]{Gabriella d'Albenzio}
\author[3]{Ruoyan Meng}
\author[1,4,5]{Davit Aghayan}
\author[1]{Egidijus Pelanis}
\author[6]{Rebecca Hisey}
\author[1,7]{Sarkis Drejian}
\author[1,8]{Åsmund Avdem Fretland}
\author[1,2]{Ole Jakob Elle}
\author[1,7,8]{Bjørn Edwin}
\author[1,3]{Rafael Palomar}

\affiliation[1]{organization={The Intervention Centre},
             addressline={Oslo University Hospital},
             city={Oslo},
             postcode={0327},
             country={Norway}}

\affiliation[2]{organization={Department of Informatics},
             addressline={University of Oslo},
             city={Oslo},
             postcode={0373},
             country={Norway}}

\affiliation[3]{organization={Department of Computer Science},
             addressline={Norwegian University of Science and Technology},
             city={Kingston},
             postcode={2815},
             country={Norway}}

\affiliation[4]{organization={Department of Surgery N1},
             addressline={Yerevan State Medical University after M. Heratsi},
             city={Yerevan},
             postcode={0025},
             country={Armenia}}

\affiliation[5]{organization={Department of Surgery, Ringerike Hospital},
             addressline={Vestre Viken Hospital Trust},
             city={Drammen},
             postcode={3004},
             country={Norway}}

\affiliation[6]{organization={Laboratory for Percutaneous Surgery},
             addressline={School of Computing within Queen’s University},
             city={Kingston},
             postcode={ON K7L 2N8},
             country={Canada}}

\affiliation[7]{organization={Institute of Clinical Medicine},
             addressline={University of Oslo},
             city={Oslo},
             postcode={0372},
             country={Norway}}

\affiliation[8]{organization={Department of HPB Surgery},
             addressline={Oslo University Hospital},
             city={Oslo},
             postcode={0372},
             country={Norway}}

\begin{abstract}

Objective: Surgical resection is currently the main curative treatment for primary and secondary liver cancer. Computer-assisted systems for liver resection planning rely on the definition of virtual resection using different geometric modeling techniques. This study aimed to develop a novel approach to simplify the definition of virtual resections and adapt to the diverse shapes of resections resulting from parenchyma-sparing resection (PSR) plans. Furthermore, we utilized this method to investigate the optimization of resection plans in the context of parenchyma-sparing approaches compared to anatomical resection (AR) plans.
Mehtods: The approach presented in this work is based on the definition of virtual resections through contours placed/extracted from the surface of the liver and spline surface approximation. As opposed to traditional methods, this naturally adheres to the surgical cutting path, as it is done during real surgery, and provides curved resection plans. Furthermore, this method can be combined with existing surface deformation techniques for increased flexibility. Both resected specimens and remnant liver volumetry derived from surgical plans performed by surgeons using the presented approach are recorded and compared. The dataset uses 14 real clinical cases from the OSLO-COMET study. 
Results: Quantitative assessment has revealed significant differences between plans corresponding to PSR and AR. PSR showed lower resected volume ($32.71 \pm 13.80$ $ml$) than AR ($249.53 \pm 135.23$ $ml$) ($p <0.0001$) and higher Remnant Volume ($1922.77 \pm 442.86$ $ml$) than AR ($1716.87 \pm 403.00$ $ml$) ($p <0.0001$). The Remnant Percentage for PSR ($98.16 \pm 0.81\%$) is significantly higher than AR ($87.40 \pm 6.49\%$) ($p <0.0001$). 
Conclusion: The proposed approach is able to define virtual resections accommodating a wide variety of resections (i.e., PSR and AR). Careful surgical planning using virtual resections can optimize the resection strategy.
Significance: This study presents a novel computer-aided planning system for liver surgery, demonstrating its efficacy and flexibility for definition of virtual resections. Virtual surgery planning can be used for optimization of resection strategies leading to increased preservation of healthy tissue. 

\end{abstract}

\begin{keyword}
Liver Resection \sep Surgical Planning \sep  Spline Theory \sep Parenchyma-Sparing.
\end{keyword}

\end{frontmatter}


\section{Introduction}
\label{sec:introduction}

Liver resection, defined as the surgical removal of liver tumors and surrounding tissue, is a widely applied treatment for cancer in the liver and the most important curative approach for various liver malignancies. \cite{pulitano2011liver, kaibori2017comparison, aghayan2018laparoscopic, ratti2018laparoscopic,fretland2019long, garbarino2020laparoscopic, petrowsky2020modern}.The complexity of this surgical procedure is attributed to the complexity of the organ itself (vast networks of blood vessels and bile ducts and different functional areas) and the nature of the disease (the tumor's location, shape, size, and number are very variable). This requires liver surgeons to have a high level of expertise, an in-depth understanding of liver anatomy, as well as a meticulous resection plan \cite{pelanis2020use}.

Broadly speaking, resections can be classified into two types: anatomical and non-anatomical resections. Anatomical resections (AR) involve the removal of continuous liver segments containing the tumors, as defined by the Couinaud classification system. Moreover, according to the \emph{Terminology Committee of the International Hepato-Pancreato-Biliary Association} \cite{strasberg2007terminology}, AR can be further classified as monosegmentectomy, bisegmentectomy, right hepatectomy, or left hepatectomy, based on the specific segments removed.
Non-anatomical liver resections, characterized by limited excision of liver tissue without adherence to segmental anatomy's drainage and blood supply, are preferred in colorectal liver metastases (CLM) for their capacity to preserve parenchymal volume and reduce postoperative complications \cite{kingham2015hepatic, aghayan2023laparoscopic}.

Liver resection can be performed under either open or minimally invasive approaches \cite{fretland2018laparoscopic}. The latter has progressed significantly in recent years, and offers numerous benefits over open surgery, including reduced complication rates, pain, and recovery time, and improved postoperative health-related quality of life \cite{agha2003does}. The laparoscopic approach to perform parenchyma-sparing liver resection is increasingly used, making these complex procedures maximally minimally invasive \cite{kazaryan2019laparoscopic, fretland2019quality, kalil2019laparoscopic,torzilli2020parenchyma}.

In recent years, computer-aided pre-operative planning systems have been developed to assist surgeons in preparing and guiding liver surgery. These systems use medical imaging data and computational algorithms to generate 3D models\textemdash generated from Computed Tomography (CT) and/or Magnetic Resonance Imaging (MRI) of liver and surrounding anatomy\textemdash and virtual resections, allowing surgeons to model different surgical scenarios and evaluate the potential outcomes \cite{yeo2018utility, felli2023paradigm}. While the amount of information available to plan liver resections is extensive (medical images and derived segmentations), current liver resection planning systems still use relatively simple approaches. Patient-specific 3D geometric models have been shown to enhance surgeons' ability to visualize the liver and its underlying vascular structures \cite{yeo2018utility}. However, to fully realize the benefits of this technology, there is a need for more widely available virtual resection planning systems based on 3D models. Such systems would enable the investigation of previously inaccessible areas that cannot be adequately captured by 2D modalities \cite{mise2013virtual}.

Computer systems provide an essential platform for pre-operative planning, enabling surgeons to make well-informed decisions and minimize the likelihood of complications \cite{lang2005impact, alirr2020survey}. The importance of a multidisciplinary approach in patient assessment cannot be overstated, requiring the collaborative efforts of oncologists, hepatobiliary surgeons, biomedical engineers, and radiologists to ensure comprehensive patient care. The debate regarding the preference for anatomical versus non-anatomical (atypical) resections in oncologic liver surgery underscores the need for planning software capable of accommodating both strategies \cite{rengers2023surgery, maher2017management, germani2022management}. However, the development and validation of effective computer-aided pre-operative planning tools for liver surgery present significant technical and clinical challenges, highlighting its importance as a key area for continued research.

This paper presents a novel computer-aided pre-operative planning system tailored for parenchyma-sparing liver resections. The system leverages patient-specific 3D models, incorporating detailed representations of the liver, tumors, and vascular structures, to provide a comprehensive visual aid for strategic surgical planning.  Our approach presents a novel method for delineating the demarcation line, which is crucial for guiding liver resections \cite{felli2020demarcation}. Additionally, we adapt deformable surfaces to the user-defined contours on the liver's surface, offering great flexibility for planning PSR, which often requires highly curved resections to spare functional liver parenchyma. We compare the surgical planning outcomes and volume estimations of PSR and AR using the proposed method and the liver segments classification method proposed in our previous work \cite{d2023patient}.  Furthermore, we have made our implementation available through the open-source software 3DSlicer \cite{kikinis20133d}, ensuring accessibility for the wider medical and research community.

\section{Related Work}
\label{sec:related}
Pre-operative liver surgery planning systems employ virtual resections as a core design concept. The specification of a virtual resection can be used for quantitative analysis regarding the volume of the intended resection or the percentage of an organ that would be removed. The literature in this field is vast, especially when surgical planning systems for other organs are considered. Here, we limit the scope  to computer-assisted systems related to resection planning; for a survey on surface reconstruction methods, we refer to \cite{christ2017computational}.

With advancements in the spatial and temporal resolution of imaging modalities and new techniques for automated path definitions, the Konrad-Verse \emph{et al.}\cite{konrad2004virtual} focused their application on deformable cutting planes. The idea in this work is to determine the orientation and extent of a discrete cutting plane generated from the lines drawn by the user on the liver surface for designing an initial planar grid, through a principal component analysis on the point set forming the lines. This planar grid can be flexibly deformed with an adjustable sphere of influence.

The goal of Ruskó \emph{et al.} in \cite{rusko2013virtual} was to perform anatomical segment separation by drawing traces on 2D slices of CT volumes and at the same time, enhancing the process using B-spline interpolation. Clearly, by drawing in slices, a cutting surface can be specified as precisely as desired. However, this process is time-consuming if the entire resection volume should be specified because often, some 50-100 slices are involved. The tool is based on an algorithm that interpolates the user-defined traces with B-spline surface and computes a binary cutting volume that represents the different sides of the surface. The computation of the cutting volume is based on the multi-resolution triangulation of the B-spline surface. 
A more straightforward interactive method was proposed by Palomar \emph{et al.}~\cite{palomar2017novel}, where a low level of user interaction is required, making the system easy-to-use by clinicians. The novelty of this method is based on the use of Bézier surfaces, which can be deformed using a grid of control points, and distance maps as a base to compute and visualize resection margins (indicators of safety) in real-time. Yang \emph{et al.}~\cite{yang2018dr}, in their surgical computer-aided system for living donor liver transplantation treatment, proposed a sphere-based virtual resection method used to divide the liver into left and right lobes. This division was performed in the axial view of the overlaid image slices, so a circle is generated by the section of the sphere over the slice. Their method, however, formed the cutting plane based on the large cutting sphere and is explicitly designed for graft resections in living donor liver transplantation.

More recently, Chheang \emph{et al.}~\cite{chheang2021collaborative} proposed a comprehensive approach to liver resection planning using virtual reality (VR). They developed a collaborative VR environment specifically designed for liver surgery planning, incorporating features such as virtual resections, risk map visualization, and volume estimation. The initial stage involves visualizing the risk map on vascular structures and projecting tumor contours on 2D slices. Resection surfaces can be defined on the organ parenchyma surface, and both 2D and 3D modifications of the virtual resection can be made with real-time risk map visualization. The authors emphasized the importance of addressing network low-latency and reliability requirements to ensure high-quality remote collaboration and accommodate user-specific precision and spatial coordination limitations. Furthermore, the same authors conducted a preliminary investigation in \cite{chheang2023virtual} to assess the usability of the Bézier surface interaction in a collaborative VR environment, as compared to a free deformation approach. The findings revealed that while some participants expressed the need for additional training with the Bézier surface interaction, they acknowledged that familiarity with this technique could potentially yield superior outcomes and facilitate faster deformation, thereby enhancing the potential for improvements. In contrast, the deformable cutting plane, although straightforward to learn, offers limited prospects for refinement or incorporating supplementary features.

\begin{figure*}[ht]
\centerline{\includegraphics[width=1\columnwidth]{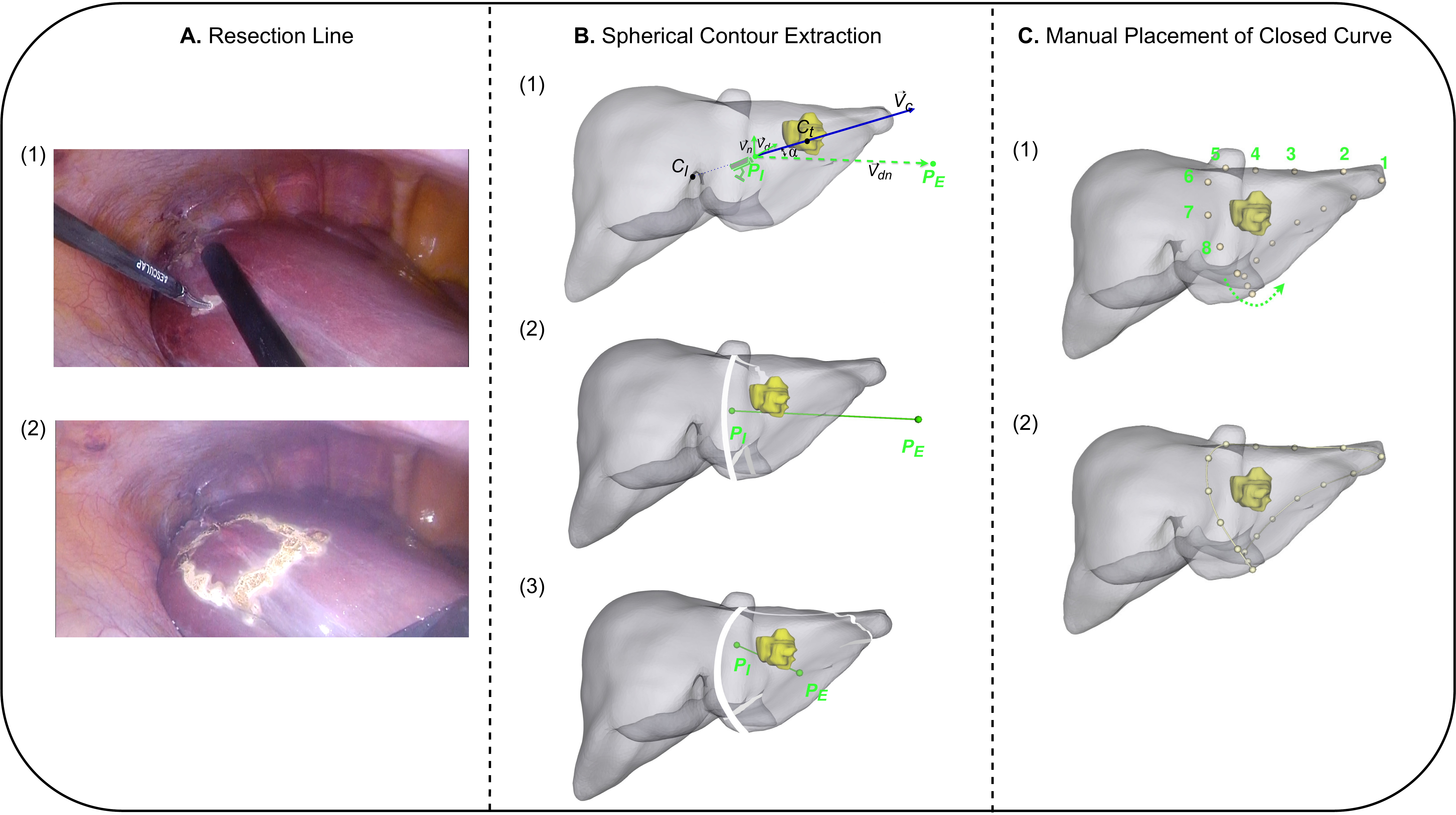}}
\caption{Visualization of the surgical resection line (A) alongside pre-operative representations using contour definition modeled with Spherical Contour (B) and Markups Closed Curve (C).}
\label{fig1}
\end{figure*}

\section{Method}

We present a specialized system for planning patient-specific PSR. First, 3D models of the liver, vessels, territories, and tumors are created using segmented CT images of the patient. Next, a 3D contour is defined on the liver surface, in alignment with the surgeon's demarcation line during surgery. Finally, a virtual resection is generated based on the contour, enabling the surgeon to make modifications to the resection plan. The implementation of this system was facilitated by the open-source software 3D Slicer \cite{kikinis20133d} and made publicly available through the Slicer-Liver extension \cite{slicerliver}. 

\subsection{Contour Definition}
During PSR, surgeons aim to minimize the removal of healthy liver tissue. They utilize a specific guideline, referred to as a \emph{resection line}, In our method, we initiate the process by replicating this demarcation process line through user-defined contours on the 3D liver surface. In this work, and as shown in Fig \ref{fig1}, we propose two different approaches to define the contour: 
\emph{Spherical Contour Extraction} and \emph{Manual Placement of Closed Curve}.

\subsubsection{Spherical Contour Extraction}
In this work, we will use the term \emph{resection contour} to denote the locus of points in the liver surface which intersect with an arbitrary surface (virtual resection). While in Palomar \emph{et. al}~\cite{palomar2017novel} a resection contour is derived from the intersection of the liver with a planar surface (e.g., cross-section), in this work we use surfaces derived from spheres, which intersection with the liver surface provides a good foundation for curved resections. In addition to the 3D models of the liver, vessel territories, and tumor, a line with two reference points (external and internal), denoted as $\mathbf{P}_{E} \in \mathbb{R}^3$ and $\mathbf{P}_{I} \in \mathbb{R}^3$, is displayed. This line, traversing the surface of the liver, is used to define the initial resection outline. The initial position of the two reference points, $\mathbf{P}_{I}$ and $\mathbf{P}_{E}$, which are utilized to guide the liver resection surgery planning process, are determined using a combination of the centers of mass of the tumor and parenchyma models, as well as the current camera view observing the 3D scene.
Specifically, $\mathbf{P}_{I}$ is positioned at the mean of the centers of mass, $\mathbf{C}_t$ and $\mathbf{C}_l$, of the 3D tumor and liver models respectively. Conversely, $\mathbf{P}_E$ is located using the camera observing the scene. This involves calculating the cross-product, $\overrightarrow{\mathbf{V}}_{nd}$, between the camera's vertical axis (up vector) and the direction vector from the camera position to the focal point, $\overrightarrow{\mathbf{V}}_{n}$ and $\overrightarrow{\mathbf{V}}_{d}$, and determining the liver surface extent, $l$. The angle $\alpha$ between the computed vector, $\overrightarrow{\mathbf{V}}_{c}$ (between tumor and liver centers of mass), and the cross-product vector is then calculated. Subsequently, $\mathbf{P}_{E}$ is positioned along one of the vectors, depending on $\alpha$. Depending on the value of $\alpha$, the external point $\mathbf{P}_{E}$ is determined by adjusting its position relative to the tumor center. If $\alpha$ is equal to or exceeds $\pi/2$ radians, $\mathbf{P}_{E}$ is placed in the direction of $-\overrightarrow{\mathbf{V}}_{nd}$, shifting by the length of the bounding box along the axis along the sagittal plane. Conversely, when $\alpha$ is less than $\pi/2$ radians, $\mathbf{P}_{E}$ is positioned in the direction of $\overrightarrow{\mathbf{V}}_{nd}$, also translating by the length of the bounding box on the sagittal plane axis. This choice is made to ensure proper alignment of the external point with the desired direction from the tumor center.
The spherical contour extraction method proposed has the ability to model precisely the resection contour in the context of PSR. Specifically, given a set ${\mathbf{P}_i}$ with ${i=0,..,m} \in \mathbb{R}^3$, which defines the spatial coordinates of the 3D model of the liver, the method involves the identification of a subset of points that closely align to a spherical surface centred around a reference point $\textbf{P}_E$. 
The method unfolds in three steps: \emph{Point Acquisition}, \emph{Distance Calculation} and \emph{Contour Extraction} as outlined in Algorithm \ref{alg:resection_contour}.
\begin{algorithm}
\label{alg:resection_contour}
\caption{Liver Resection Spherical Contour Extraction}
\begin{algorithmic}[1]
\STATE \textbf{Input:} reference points $P_I$, $P_E$; liver surface $LiverS$; contour thickness $k$
\STATE \textbf{Output:} Spherical Contour Points

\STATE \textbf{Initialization:} 
\STATE Spherical Contour Points $\leftarrow 0$

\STATE \underline{\emph{Point Acquisition:}}
\FORALL{points $P_i$ in $LiverS$}
    \STATE Store the geometry of $LiverS$ in a point set.
\ENDFOR

\STATE \underline{\emph{Distance Calculation:}}
\FOR{each point $P_i$ in point set}
    \STATE Calculate $d_i = \|P_i - P_E\|_2$.
\ENDFOR

\STATE \underline{\emph{Contour Extraction:}}
\STATE Define $d_{\text{ref}} = \|P_I - P_E\|_2$ as the reference distance.

\FOR{each $d_i$}
    \IF{$|d_i - d_{\text{ref}}| < k$}
        \STATE Add $P_i$ to Spherical Contour Points.
    \ENDIF
\ENDFOR

\STATE \textbf{Note:} $k$ defines the thickness of the contour (i.e., in our implementation, we use $k=0.07$, which corresponds to the thickness of the contour used for visualization)
\end{algorithmic}
\end{algorithm}

\subsubsection{Manual Placement of Closed Curve}
In the manual delineation of resection boundaries, the process involves defining a continuous, closed curve whith all its points lying on the surface of the liver Using this method, the surgeons carefully mark out specific areas for resection by positioning control points on a 3D model of the liver in a three-dimensional space. The interpolation between these points employs spline smoothing techniques to ensure the creation of a continuous and aesthetically coherent curve \cite{de1978practical}. For the implementation of this method in our study, we leverage the capabilities of Markups in 3D Slicer \cite{kikinis20133d} to facilitate the manual placement of control points and curve generation.

\begin{figure}[!ht]
\centerline{\includegraphics[width=1\columnwidth]{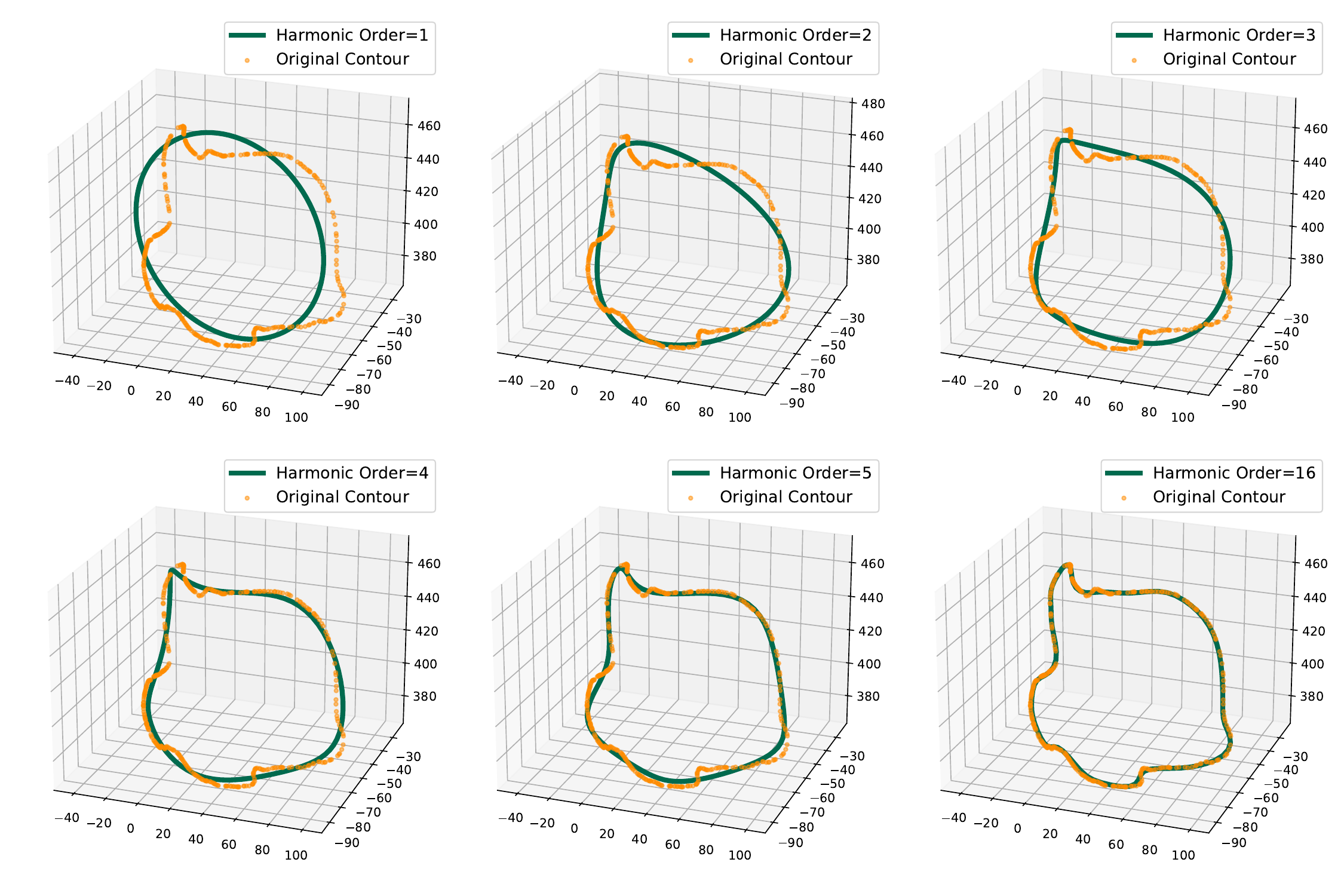}}
\caption{Visualization of Harmonics through Elliptic Fourier Analysis on a 3D Closed Contour: The initial five diagrams illustrate the reconstructions for the first five harmonics, presented in green. In contrast, the final diagram showcases the 16th harmonic's contour reconstruction at the Nyquist frequency, also displayed in green. The original closed contour is depicted in orange.}
\label{efa}
\end{figure}

\subsection{Shape Analysis-Based Contour Optimization}
Given that the demarcation line of any 3D liver model can be approximated as a closed 3D contour with $x-$, $y-$, and $z-$coordinates, we employ a custom 3D Elliptic Fourier analysis (EFA) to smooth the contour, eliminate redundant data, and generate equidistant sample 3D points representing the contour. This refined data is crucial for subsequent stages of the algorithm as it structures the data topology for the next parameterization step in which a unique parameter value to each point in the 3D contour for use in fitting of surfaces. We extend the Kuhl and Giardina EFA \cite{kuhl1982elliptic}, which originally works on 2D datasets, to produce 3D curves. Through 3D EFA, the liver's resection outline can be reconstructed using harmonic equations, enabling contour representation using a reduced set of parameters. The number of Fourier coefficients is determined by the product of the number of harmonics ($k$) and the number of $x$-, $y$-, and $z$-projections in $\mathbb{R}^3$. Each harmonic equation comprises two coefficients per dimension, resulting in a total of $6 \times k$ coefficients.
A 3D closed curve parameterized by $0 < t < 2\pi$ can then be expressed as a weighted sum of Fourier basis functions:

\begin{equation}
    \begin{bmatrix}
        x(t)\\ y(t)\\ z(t)
\end{bmatrix} 
=   \begin{bmatrix} 
        \alpha_0\\\gamma_0\\\varepsilon_0
\end{bmatrix} + \sum_{k=1}^{K}
    \begin{bmatrix}
        \alpha_k & \beta_k\\
        \gamma_k & \delta_k\\
        \varepsilon_k & \zeta_k\\
\end{bmatrix}
    \begin{bmatrix}
        \cos{kt}\\
        \sin{kt}
    \end{bmatrix}
\end{equation}
where $k$ is the maximum number of harmonics used. Let $\mathbf{f}(t) = \begin{bmatrix} x(t) & y(t) & z(t)\end{bmatrix}^T$ be the vector-valued function and $\mathbf{g}_k(t) = \begin{bmatrix}\cos{kt} & \sin{kt}\end{bmatrix}^T$ be the vector-valued function for each $k$.
For a 3D closed curve, the coefficients are given by the following expressions: 
\begin{equation}
\begin{aligned}
    \begin{bmatrix}
        \alpha_0\\ \gamma_0\\ \varepsilon_0
    \end{bmatrix}
    &= \frac{1}{2\pi} \sum_{j=0}^{m-1} \mathbf{f}(t) \, dt, &
    \begin{bmatrix}
        \alpha_k\\ \beta_k\\ \gamma_k\\ \delta_k\\ \varepsilon_k\\ \zeta_k
    \end{bmatrix}
    &= \frac{1}{\pi} \sum_{j=0}^{m-1} \mathbf{f}(t) \cdot \mathbf{g}_k(t) \, dt
\end{aligned}
\end{equation}

in which $tj=2\pi j$, where $j$ assumes the values $0,1,\dots,m-1$, and $m$ represents the number of points utilized to reconstruct the contour.
The total number of harmonics computable for any contour is half the total number of contour coordinates, adhering to the Nyquist frequency principle. Fourier methods enable efficient shape reconstruction by considering the first few harmonics, capturing the primary shape attributes in a concise format.

The required number of harmonics for a desired precision level is estimated from the average Fourier Power Spectrum ($F_p$) \cite{crampton1995elliptic, costa2009quantitative}:

\begin{equation}
F_p(t)= \frac{\sum_{k=1}^{K}\alpha_{k}^2 + \beta_{k}^2 + \gamma_{k}^2 + \delta_{k}^2 + \varepsilon_{k}^2 + \zeta_{k}^2}{3}
\end{equation}

Consequently, the Fourier series can be truncated at $K$ to retain 99.99\% of the average total power of the contour.

\subsection{Contour Filling}
This section outlines the approach for generating interior points following the reconstruction of a 3D contour using EFA. 
We aim to prepare 3D grid data for surface fitting using cubic B-spline interpolation. Cubic splines are particularly advantageous due to their balance between computational efficiency and the ability to fit data smoothly. They are defined by piecewise cubic polynomials that not only interpolate the data points but also maintain $C^2$ continuity, ensuring smooth transitions between segments.
The choice of cubic splines is driven by their property of energy minimization. This feature ensures that among all feasible $C^2$ interpolating functions, the cubic spline possesses the minimal energy, resulting in an optimally smooth curve with diminished oscillations \cite{WOLBERG2002145}. Given a sequence of contour data points $\textbf{D} = (D_0, D_1,...D_m)^T \in \mathbb{R}^{3}$,  our goal is to create a smooth curve that passes through these points in a specific order, specifically connecting points at even indices with those at odd indices. This is achieved through a B-Spline curve, which provides a continuous interpolation between these points. To accomplish this, we use a cubic B-Spline curve that provides a smooth and continuous interpolation between these points.
A B-spline curve of degree $d=3\in \mathbb{N}$  is defined as:
\begin{equation}
	\textbf{Q}(t)=\sum_{i=1}^{m}\textbf{c}_{i}B_{i,3}(t)
\end{equation}
where $\textbf{c} = \{c_i\}_{i=1}^{m} \in \mathbb{R}^{3}$ are the control points of \textbf{Q} which define the control polygon passing through those points and $\textbf{t} = \{t_i\}_{i=1}^{m+3+1} \in \mathbb{R}$ is the knot vector with non-decreasing elements. The basis functions $B_{i,3}$ are defined recursively as follows:
\begin{equation}
\left\{
\begin{aligned}
 B_{i,0}(t) & = 1 \quad \text{if } t_{i}\le t \le t_{t_1+1} \\ 
 B_{i,0}(t) & = 0 \quad \text{otherwise} 
\end{aligned} \right.
\end{equation}
\begin{equation}
\begin{aligned}
B_{i,3}(t)=& \frac{t-t_i}{t_{i+3}-t_i}B_{i,2}(t) + \frac{t_{i+4}-t}{t_{i+4}-t_{i+1}}B_{i+1,2}(t)
\end{aligned}
\end{equation}
Our specific problem of interpolating a series of 3D points using a spline is essentially a univariate interpolation problem, and the B-spline curve can be parametrically represented as a 3D curve: 
\begin{equation}
\textbf{Q}(u) = (x (u), y (u), z ( u))
\end{equation}
Here, $u$ is the parameter ranging from 0 to 1.
When working with curves, the selection of knots by specifying the parameter values corresponding to the data points is crucial. This choice impacts the shape and approximation order of the curve \cite{floater2006parameterization}. In this study, we employ a chord-length parametrization as it provides full approximation order for cubic interpolation \cite{floater2006parameterization}. Consequently, we define the knots using the knot averaging method \cite{de1978practical} as described in \cite{ammad2019cubic}.
The objective of computing B-spline curve interpolation is to determine a set of control points $\mathbf{c}_i$ given a fixed number of control points $m$, a degree $d$, and a knot vector $\mathbf{t}$ with which the B-spline curve interpolates a set of curve points at specific parameter value $u_i$ for $i=1,\dots,m$. 
This is achieved by solving the following system of linear equations:
\begin{equation}
 \mathbf{Q}(u) = \sum_{i=1}^{m}\mathbf{c}_{i}B_{i,d}(u_i)
\end{equation}
where the control points $\mathbf{c}_i$ are the unknowns. 

The control points can be obtained by solving this system of equations using Gauss elimination. Subsequently, these control points can be used to construct a curve that accurately passes through each data point.

\begin{figure*}[ht]
\centerline{\includegraphics[width=1\linewidth]{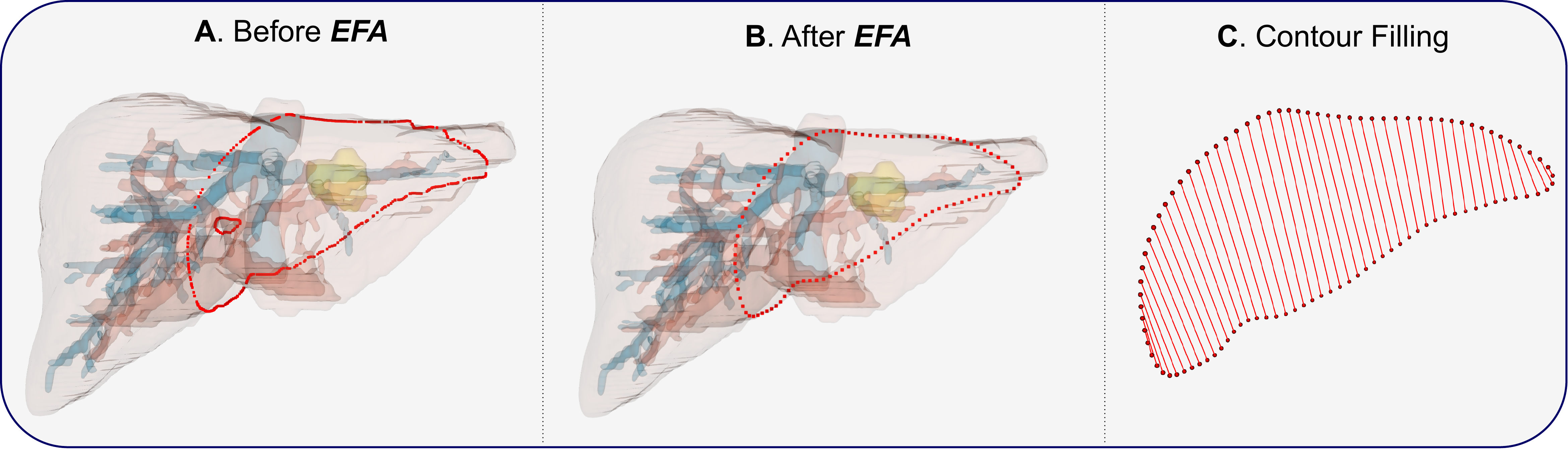}}
\caption{The figure depicts the progressive stages of our hepatic resection planning technique culminating in contour filling: (A) Initial contour delineation is performed using spherical extraction prior to the application of Elliptic Fourier Analysis (EFA); this stage includes artifacts (isolated regions generated by the liver's anatomical variations) that are subsequently eradicated and the contour refined through EFA as shown in (B); finally, (C) illustrates the implementation of the contour filling strategy.}
\label{filling}
\end{figure*}

\subsection{Surface Approximation}
After obtaining the spline curves, we can now create a 3D grid representation of the resection surface. To form the resection surface from the grid, we expand our method to include general spline tensor-product surfaces. However, we have decided to focus our experiments on bicubic tensor-product Bézier surfaces given their demonstrated advantages in planning, as discussed in \cite{palomar2017novel}. Bicubic tensor-product Bézier surfaces, a subset of mathematical splines, offer an optimal framework for modeling polynomial patches due to their inherent global smoothness and the property of remaining confined within the convex hull formed by their control points. For a comprehensive explanation and validation of these characteristics, readers are referred to \cite{cohen2001geometric}.
A bicubic tensor-product Bézier surface in $\mathbb{R}^{3}$ is a parametric polynomial surface $\mathbf{P} : D \rightarrow \mathbb{R}^{3}$ of degree $3\times 3$, given by the formula:
\begin{equation}
    \mathbf{P}(u,v)=\sum_{i=0}^{3}\sum_{j=0}^{3}\mathbf{c}_{i,j}S^{3}_{i}(u)S^{3}_{j}(v)\mathrm{,} \quad \mathbf{c}_{i,j}\in \mathbb{R}^3
\end{equation}
where $u$ and $v$ are parametric coordinates defined within the rectangle domain  $D=[0, 1] \times [0,1]$.
The vector $\mathbf{c}_{i,j}$ represent the $4\times4$ control points, which form the control net of $\mathbf{P}(u,v)$, furthermore $S^{3}_{i}(u)$ and $S^{3}_{j}(v)$ are the Bernstein basis polynomial defined as:
\begin{equation}
    S^{3}_{l}(p) = \binom{3}{l}p^l(1-p)^{3-l}, \quad p\in[0,1]
\end{equation}

Here, $p$ represents a generic parameter that can either be $u$ or $v$ and $l$ is an index that can take the place of either $i$ or $j$, depending on whether $p$ is $u$ or $v$.

This problem can be solved by means of least-square approximation for 3D gridded data. For this, let $\mathbf{G}_{k_1,k_2}$  with $k_1 = 1,2, \dots, n_1$ and $k_2 = 1,2, \dots, n_2$ be the input 3D set of points arranged in a quadrilateral structure and $C_{i,j}$, $i,j\in \{0,...,3\}$ the set of unknown Bézier control points, which give the best fit in the discrete least-squares sense, this is, solving the following minimization problem for the squared errors $\mathbf{E}$:
\begin{equation}
	\mathbf{E}=\sum_{k_1=0}^{n_1} \sum_{k_2=0}^{n_2} (\mathbf{G}_{k_1,k_2} - \mathbf{P}(u_{k_1},v_{k_2}))^2
\label{eq12}
\end{equation}
Solving the least-squares minimization problem means solving the following system of equations:
\begin{equation}
	[G] = [C][S]
\end{equation}
where, $[G]$ is the vectorization of the 3D data points, $[C]$ corresponds to the vectorization of the control points net and $[S]$ is the matrix representation of the Bernstein basis polynomials.
The algebraic solution of \eqref{eq12} is given by:
\begin{equation}
	[C]= [S]^{-1}[G]
\end{equation}

The proposed algorithm has been expanded within our Slicer-Liver extension \cite{slicerliver} to accommodate Non-Uniform Rational B-Spline (NURBS) surfaces, with uniform weights assigned to all control points (weight value of 1) \cite{d2024using}.

\subsection{Experimental Design}
 The experimental setup is based on a subset of the Oslo-CoMet trial dataset provided by The Intervention Centre at Oslo University Hospital in Norway, comprising 14 cases chosen by two surgeons to ensure diversity in tumor diameters and locations \cite{fretland2015open, fretland2018laparoscopic}. The dataset includes segmentations of liver parenchyma, vascular structures, and lesions. Segmentations were processed by the flying edges algorithm \cite{schroeder2015flying} followed by smoothing and decimation to generate 3D models of the labeled tissues.

\subsubsection{Comparative Analysis of Liver Surgery Approaches }
This study aims to assess the application of virtual resections in the pre-operative planning for liver surgeries. Our focus was on exploring the differences between two liver surgery techniques employed in the treatment of colorectal liver metastases: PSR (Parenchyma-sparing Resection), which is inclined towards atypical resections, and AR (Anatomical Resection). We compared these techniques, particularly looking at how atypical and anatomical resections differ in terms of the volume of liver tissue removed.
For PSR, we applied our proposed method, whereas for AR, we utilized a technique based on vessel centerline extraction and developed in our centre, as detailed in \cite{d2023patient}. 
Both methods are implemented using 3D Slicer \cite{kikinis20133d} and are publicly available in the Slicer-Liver extension. After generating surgical plans, we calculated the surgical planning outcome metrics for patients undergoing hepatic surgeries with either AR or PSR and provided visualizations of the plans using the same software \cite{kikinis20133d}. 

\subsubsection{Dataset and Statistics}
We conducted our assessment using data from the Oslo-CoMET study \cite{fretland2015open, fretland2018laparoscopic}, comprising over 200 medical images of patients with segmented liver, tumor, and vessel territories. From this dataset, two surgeons selected 14 cases for PSR and AR planning based on tumor characteristics, size, and location. Patient 12 was the only case with two tumors in the same liver segment, whereas the other patients had a single liver tumor metastasis. The quality of portal and hepatic vein segmentation played a crucial role in case selection.
Before the experiments, surgeons familiarized with a graphical user interface and underwent training. The experiments involved planning the 14 cases using the two methods. The most experienced surgeon (37 years of clinical experience) performed PSR planning using our proposed method, while the other (8 years of clinical experience) performed AR resection planning using the centerline extraction-based method \cite{d2023patient}. 
For the statistic analysis, the Shapiro-Wilk test \cite{shapiro65} is used to evaluate the normality of data. Pairwise differences between AR and PSR methods are analyzed using the paired Student’s t-test \cite{student1908probable} or Wilcoxon signed-rank test. The effects resulting from these comparisons are reported at a significance level of 0.05. The statistical analyses were performed using the R software \cite{rsoftware}.

\begin{table*}[ht]
\centering
\caption{Quantitative Assessment of Liver Resection Performance Metrics: Comparative Analysis of Point-to-Surface Distance, Resected Volume, Remnant Volume, and Remnant Percentage in AR and PSR. Average values and standard deviations are presented for each metric.}
\label{tab:my-table}
\resizebox{\textwidth}{!}{
\begin{threeparttable}
\begin{tabular}{@{}ccccccccccc@{}}
\toprule
\textit{\textbf{Case}} & \textit{\textbf{Tumor Location}} & \textit{\textbf{\begin{tabular}[c]{@{}c@{}}Tumor Diameter \\ (mm)\end{tabular}}} & \multicolumn{2}{c}{\textbf{\begin{tabular}[c]{@{}c@{}}Safety Margin \\ (mm)\end{tabular}}} & \multicolumn{2}{c}{\textbf{\begin{tabular}[c]{@{}c@{}}Resected Volume \\ (ml)\end{tabular}}} & \multicolumn{2}{c}{\textbf{\begin{tabular}[c]{@{}c@{}}Remnant Volume \\ (ml)\end{tabular}}} & \multicolumn{2}{c}{\textbf{\begin{tabular}[c]{@{}c@{}}Remnant Percentage \\ (\%)\end{tabular}}} \\ \midrule
\textit{\textbf{}} & \textit{\textbf{}} & \textit{\textbf{}} & \textit{\textbf{AR}} & \textit{\textbf{PSR}} & \textit{\textbf{AR}} & \textit{\textbf{PSR}} & \textit{\textbf{AR}} & \textit{\textbf{PSR}} & \textit{\textbf{AR}} & \textit{\textbf{PSR}} \\ \cmidrule(l){4-11} 
1 & S6 & 9.88 & 5.96 & 1.60 & 287.46 & 48.29 & 1435.88 & 1675.05 & 83.32 & 97.20 \\
2 & S4B & 10.85 & 7.93 & 2.56 & 51.69 & 14.02 & 1736.24 & 1766.80 & 97.11 & 99.21 \\
3 & S2/3 & 25.74 & 4.05 & 0.09 & 289.01 & 26.91 & 1148.58 & 1410.68 & 79.90 & 98.13 \\
4 & S2/3 & 21.50 & 0.94 & 2.46 & 418.07 & 32.82 & 1211.66 & 1596.90 & 74.35 & 97.99 \\
5 & S4A & 23.68 & 4.24 & 1.17 & 185.98 & 20.82 & 1739.85 & 1905.01 & 81.70 & 98.92 \\
6 & S7 & 25.14 & 24.28 & 9.96 & 352.37 & 36.54 & 1573.46 & 1889.29 & 90.34 & 98.10 \\
7 & S4A & 25.22 & 4.74 & 2.24 & 86.07 & 38.69 & 1537.80 & 1585.18 & 94.7 & 97.62 \\
8 & S2/3 & 27.68 & 1.48 & 0.86 & 68.12 & 26.01 & 1280.04 & 1322.15 & 94.95 & 98.07 \\
9 & S6 & 35.72 & 1.31 & 3.14 & 137.00 & 48.05 & 1863.77 & 1952.72 & 93.15 & 97.6 \\
10 & S3 & 40.22 & 3.15 & 2.29 & 275.49 & 46.87 & 2347.89 & 2576.51 & 89.5 & 98.21 \\
11 & S8 & 11.01 & 18.64 & 1.70 & 293.37 & 22.03 & 2045.10 & 2292.09 & 87.46 & 99.05 \\
\multirow{2}{*}{ 12*} & \multirow{2}{*}{S8} & 27.93 & 3.30 & 2.82 & \multirow{2}{*}{460.02} & \multirow{2}{*}{47.06} & \multirow{2}{*}{2194.08} & \multirow{2}{*}{2563.99} & \multirow{2}{*}{82.67} & \multirow{2}{*}{98.2} \\
 &  & 13.62 & 1.81 & 0.57 &  &  &  &  &  &  \\
13 & S4A & 15.44 & 9.30 & 3.71 & 195.41 & 11.01 & 2163.46 & 2330.49 & 91.72 & 99.53 \\
14 & S6 & 27.33 & 6.17 & 3.92 & 182.90 & 53.41 & 1281.23 & 1410.72 & 87.51 & 96.35 \\ \midrule
Average &  & 22.73 & 6.49 & 2.60 & 249.53 & 32.71 & 1716.87 & 1922.77 & 87.40 & 98.16 \\
SD &  & 9.07 & 6.63 & 2.31 & 135.23 & 13.80 & 403.00 & 442.86 & 6.49 & 0.81 \\ \bottomrule
\end{tabular}
\begin{tablenotes}
\footnotesize
\item[*] \emph{In Case 12, two distinct tumors were identified within segment S8, both of which were resected simultaneously.}
\end{tablenotes}
\end{threeparttable}
}
\end{table*}

\subsection{Evaluation Metrics}
To compare the different methods, we established the criteria and corresponding objective evaluation metrics as follows:

\paragraph{Volume Assessment (Resected and Remnant)} Our procedure for calculating the resected volume is comprised of three stages. Based on the control points of the current Bézier resection surface, a high-resolution Bézier resection surface is first generated. Next, all points on this surface are mapped onto a copy of the original liver parenchyma volume image used to generate the patient liver model, where the background lable value $l_b = 0$ and the liver parenchyma label value $l_p = 1$. Mapping the high-resolution surface involves finding the voxels to which each surface point belongs, resulting in a thickened resection surface to maintain its continuity and create a clear boundary between the resected and remnant liver volumes with a label value $l_r$ that is different from and larger than any other structure in that volume. Finally, a connected threshold region growing algorithm is applied using a seed point that is arbitrarily selected from the resected area or automatically selected from the target tumor, with a lower threshold value $l_p$ and upper threshold value $l_r -1 $.\\
To obtain accurate volume calculations, it is necessary that the resection path completely enters and exits the parenchyma. This is not only consistent with the goals of the application, but also ensures a clear separation between the resected and remnant liver volumes.
\paragraph{Safety Margin} In surgical planning, achieving a balance between preserving a significant portion of the organ and ensuring complete removal of diseased tissue is crucial. To address this, we employed the \emph{point-to-surface} distance. By doing so, we can pinpoint the nearest location on the tumor surface for each position on the resection surface. This approach contributes to evaluating how closely the resection aligns with the tumor. Let $R~=~\{\mathbf{r}_1, \mathbf{r}_2, \dots, \mathbf{r}_n\}$ be the set of $n$ points on the resection surface, and $T~=~\{\mathbf{t}_1, \mathbf{t}_2, \dots, \mathbf{t}_m\}$ be the set of $m$ points on the surface of the tumor. The point-to-surface distance between each point $\boldsymbol{r}_i$ in $R$ and each point $\boldsymbol{t}_j$ in $T$ is defined as:
    \begin{equation}
    d(\mathbf{r}_i, \mathbf{t}_j) = |\mathbf{r}_i - \mathbf{p}_j|        
    \end{equation}
    where $\mathbf{p}_j$ is the closest point on the surface of the tumor to point $\mathbf{r}_i$. The minimum distance for each point $\mathbf{r}_i$ is then defined as:
    \begin{equation}
    d(\mathbf{r}_i, T) = \min{d(\mathbf{r}_i, \mathbf{t}_j) \mid \mathbf{t}_j \in T}\\       
    \end{equation}
\paragraph{Remnant Percentage}The adequacy of the remaining liver volume is contingent upon the patient's health status and existing liver conditions. Practical thresholds, grounded in empirical values, are utilized to establish the minimum volume required to prevent postoperative organ failure. It is generally recommended that the remnant volume constitutes at least $21\%$ of the estimated total liver volume. However, in cases of liver impairment due to factors like chemotherapy, steatosis, hepatitis, or cirrhosis, this percentage may need to be adjusted to the range of $30-60\%$ or even $40-70\%$ \cite{pawlik2008expanding}. Assessing the remnant percentage of the liver volume is crucial for gauging the safety and feasibility of liver resection surgery. This metric represents the proportion of the total liver volume that remains after the removal of the diseased segment. The remnant volume percentage $R_{\%}$ is determined by dividing the virtual pre-operative remnant volume $V_{r}$ by the virtual pre-operative total volume $V_{total}$:
    \begin{equation} 
        R_{\%} = \frac{V_{r}}{V_{total}} \times 100 
    \end{equation}

\section{Results}
\subsection{Quantitative Assessment of Liver Resection Performance Metrics}

Table \ref{tab:my-table} provides a comprehensive evaluation of liver resection performance metrics, comparing Anatomical Resection (AR) and Parenchyma-Sparing Resection (PSR). The evaluation metrics include Tumor Volume, Safety Margin, Resected Volume, Remnant Volume, and Remnant Percentage. Average values and standard deviations are reported for each metric.
The average tumor volume across all cases was $4.16 \pm 3.21$ $ml$. 

A statistically significant difference in the Safety Margin was observed between PSR  and AR as determined by the Wilcoxon signed-rank test ($V = 109$, $p = 0.003$).

In terms of Resected Volume, AR and PSR showed significant differences, with AR having an average volume of $249.53 \pm 135.23$ $ml$, which is significantly higher than PSR's $32.71 \pm 13.80$ ml ($t(13) = 5.8571$, $p-value = 5.619e-05$).

The Remnant Volume also demonstrated a notable difference, with AR and PSR averaging $1716.87 \pm 403.00$ $ml$ and $1922.77 \pm 442.86$ respectively, signifying a higher preservation of liver volume in PSR ($t(13) = -6.0203$, $p-value = 4.302e-05$).

\begin{figure}[ht]
\centerline{\includegraphics[width=1\columnwidth]{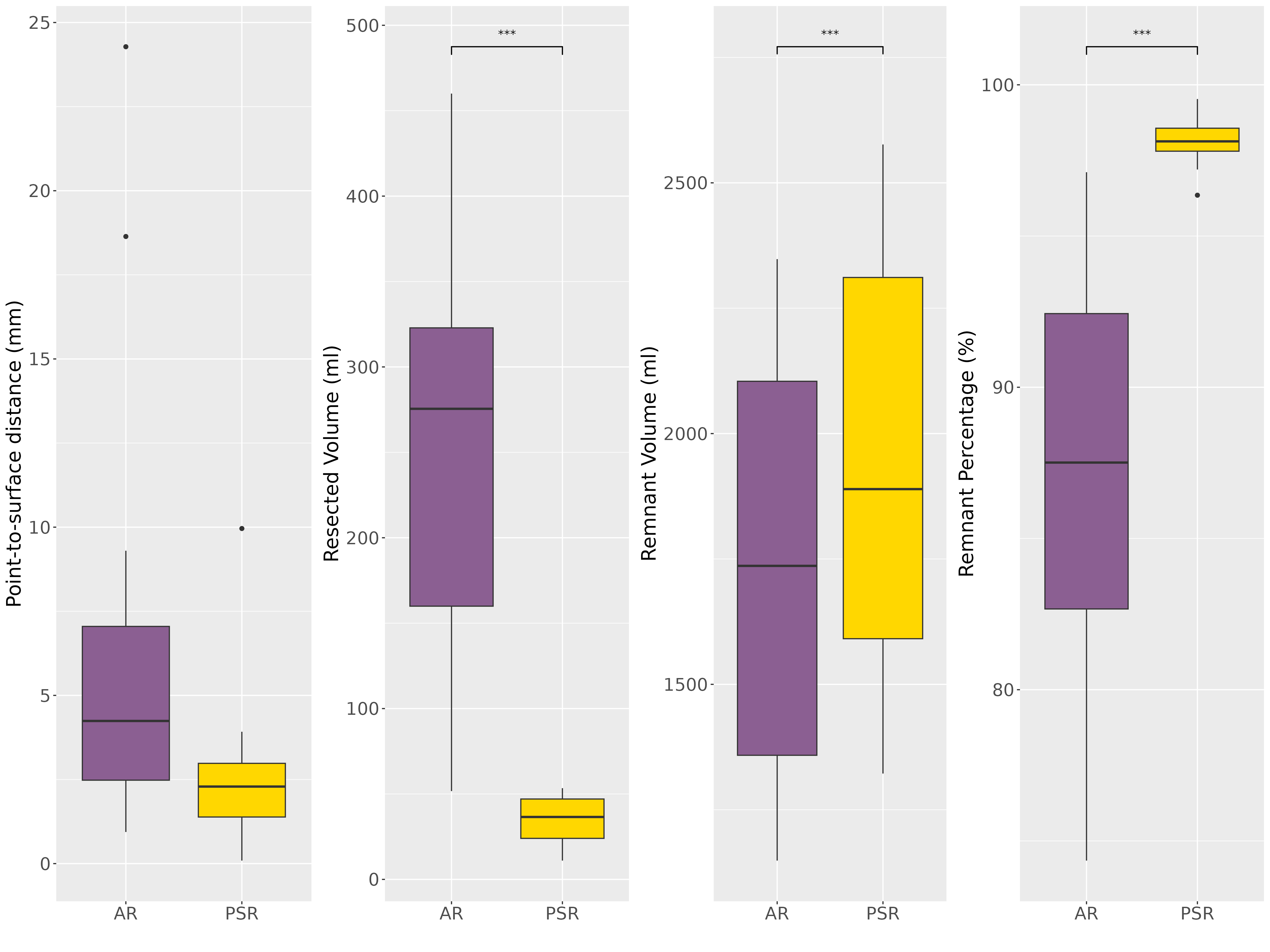}}
\caption{Comparison of Surgical Outcome Metrics in Patients Undergoing Parenchyma Sparing Resection (PSR) and Anatomical Resection (AR).}
\label{fig2}
\end{figure}

The Remnant Percentage showed an average of $87.40 \pm 6.49\%$) for AR and $98.16 \pm 0.81\%$ for PSR. The t-test for Remnant Percentage exhibited a highly significant difference ($t(13) = -6.20$, $p-value = 3.23e-05$). In addition the boxplot in Figure \ref{fig2} indicates a more condensed distribution of PSR compared to AR. This is evidenced by the tighter interquartile range of PSR, suggesting less variability in the percentage of tissue remaining post-resection.

\begin{figure*}[!ht]
\centerline{\includegraphics[width=\linewidth]{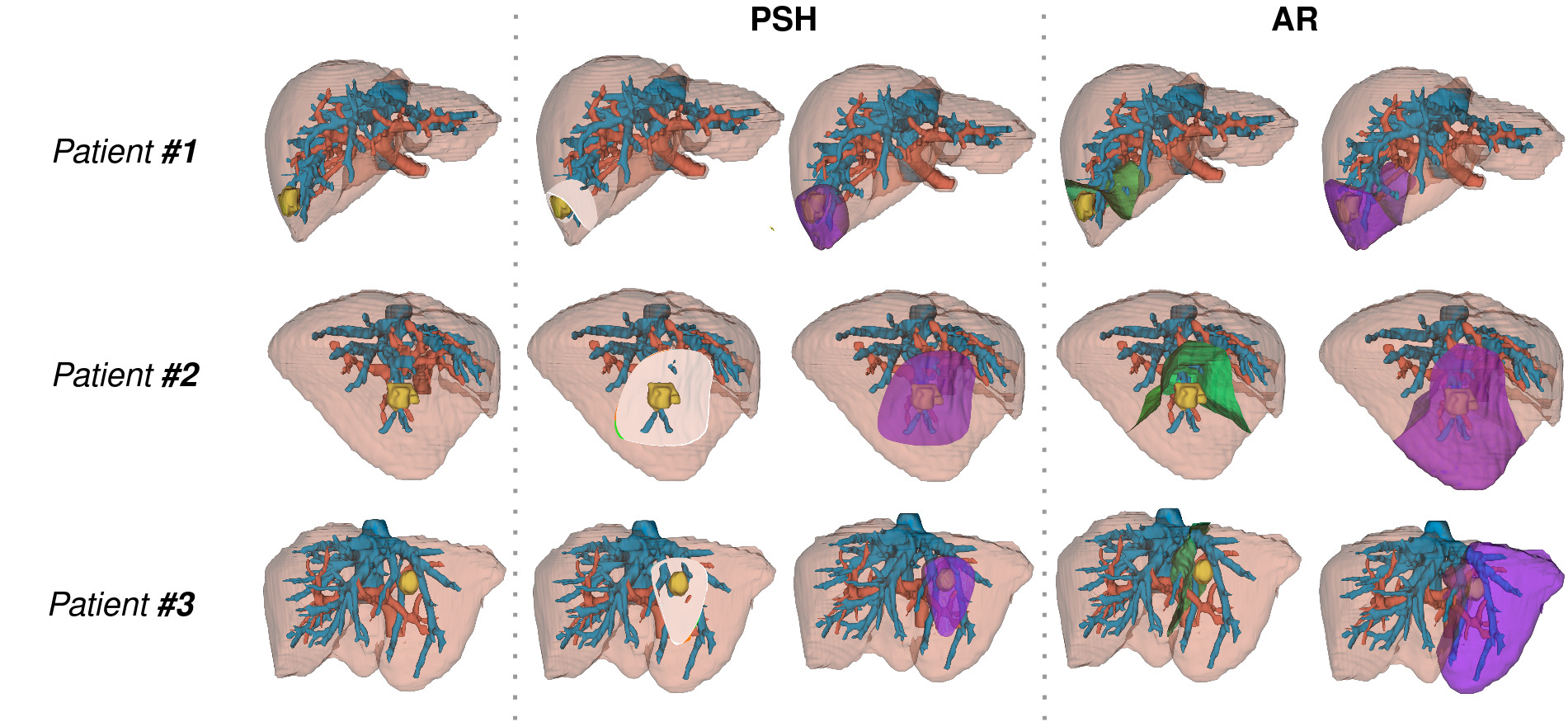}}
\caption{Comparative visualization of hepatic resection strategies across three distinct patient cases: The first column presents three-dimensional reconstructions of the liver, vascular structures, and tumors for each patient. The second column illustrates the proposed resection surface (highlighted in white) and the corresponding parenchyma volume intended for resection (shown in purple) in the context of parenchyma-sparing hepatectomy (PSR). The third column depicts the resection surface (in green) and the resected liver parenchyma volume (in purple) for anatomical resection (AR).}
\label{fig3}
\end{figure*}

\section{Discussion}
The primary goal of this work is to explore the benefits of integrating virtual resection models into the pre-operative planning workflows for liver surgery. To do this, we introduce a novel computer-assisted planning system specifically designed for parenchyma-sparing liver resections. Parenchyma-sparing resections, as opposed to anatomical resections, seek to preserve as much healthy tissue as possible by not adhering to the anatomy of the liver. We conduct a comparative analysis to assess the utility of our system, focusing particularly on the evaluation metrics for liver resection planning between anatomical resections and parenchyma-sparing resections using 14 CT volumes from the OSLO-COMET dataset \cite{fretland2018laparoscopic}.

In clinical terms, optimization of liver resections (in terms of resected volume) is performed by means of parenchyma-sparing resections (as opposed to anatomical resections). With this in mind, our method seeks to provide surgeons with a 3D computer-aided system capable of generating clinically viable surgical plans, thereby facilitating informed decision-making regarding the most appropriate surgical strategy. The mathematical formulation of the resection line in our system is designed to emulate the principles traditionally employed by surgeons during the actual execution of PSR. By following these principles, the proposed method not only brings familiarity to the surgeons during planning, but also tailors the method to PSR planning.

Building upon the idea presented in \cite{demedts2010evaluation}, which emphasizes the significance of objectively quantifying the quality of various resection strategies, we have conducted a quantitative assessment of the proposed resection plans. This evaluation was carried out using specific metrics: Safety Margin, Resected Volume, Remnant Volume, and Remnant Percentage. These metrics serve as indicators of treatment plan efficacy for the surgical plans under consideration.

Our quantitative liver resection performance metrics assessment revealed notable distinctions between AR and PSR. Our findings indicate a notably larger safety margin for AR compared to PSR, due to the removal of entire liver segments. While a close resection margin is linked to an increased risk of surgical-margin recurrence, there is no consensus that a predicted surgical margin of less than $1$ $cm$ following the resection of hepatic colorectal metastases should be considered a contraindication for surgery \cite{pawlik2005effect}; this, in turn, allows for surgical plans to minimize the resected tissue. In our series,  in three cases, the resection margin was under $1$ $mm$ in the PSR group. In these cases, the surgeon chose to perform the resection with a narrow margin, intentionally avoiding the inclusion of large vessels. This strategy aimed to minimize the volume of tissue to be resected. 
Notably, Case 3, where the safety margin for PSR was recorded at a minimum of $0.09$ $mm$, resulted in a R1-vasc resection plan \cite{donadon2019r1}, which involved separating the major left intrahepatic vessels from the lesion. This example emphasizes the challenges and considerations in balancing resection adequacy with the preservation of vital liver functions.
Caution should be exercised when interpreting the clinical significance of larger margins. There is no direct correlation between these factors and improved survival rates or reduced recurrence \cite{bodingbauer2007size, margonis2015intraoperative, martinez2021impact}. This consideration is particularly important when acknowledging the vital importance of preserving healthy liver tissue.

The significantly lower resected volumes in PSR indicate its less invasive approach, aiming to preserve as much healthy liver tissue as possible, which is crucial for patients with limited hepatic reserve or multifocal liver disease. The remnant volume and percentage findings underscore PSR's ability to leave a larger volume of the liver intact, which is essential for postoperative liver function and overall patient recovery.
Furthermore, preserving healthy liver tissue increases the possibility for application of salvage procedures. It is worth noting that a significant proportion (around 50-60\%) of patients with colorectal liver metastases experience liver recurrences following liver resection. Therefore, the ability to conduct re-do liver resections for recurrences is crucial and can extend patients' survival \cite{mise2016parenchymal, barkhatov2022long}. 
Finally, our method may help to reduce the risk of ischemia in the remaining liver tissue by providing detailed information on the portal veins included in the resection area. Ischemia in the remnant liver has been linked to poorer survival outcomes \cite{yamashita2017remnant, aghayan2018laparoscopic}.

To the best of our knowledge, this study represents the first pre-operative comparison of the individual impacts of the two resection strategies, AR and PSR, utilizing a 3D planning tool. Indeed, proposing different resection plans, along with conducting a volumetric analysis of the liver using a 3D system, should be a prerequisite for potentially improving the surgery and patient safety. Our findings underscore the efficacy of PSR in preserving more healthy liver, indeed the substantial differences in Resected Volume and Remnant Percentage highlight the unique contributions of our proposed method. 

The absence of intra-operative CT scans in the OSLO-COMET dataset has been a limitation for our study as it hampers the precision of retrospective analyses concerning actual remnant volumes and virtual remnant liver volumes. Nonetheless, a new dataset is currently being compiled at Oslo University Hospital and will include intra-operative CT scans, thereby facilitating further evaluation of this approach.

In conclusion, our study contributes to the ongoing discussion on liver surgery planning by offering a detailed quantitative analysis of AR and PSR resection approaches. PSR stands out for its ability to preserve liver volume, which is pivotal for patient recovery and long-term outcomes. These findings suggest that the choice between AR and PSR should be tailored to individual patient characteristics, including tumor location, liver function, and the presence of underlying liver disease, to optimize surgical and oncological outcomes. While our methodology presents promising results, ongoing advancements in imaging technology and dataset quality will undoubtedly refine the accuracy of virtual representations in pre-operative planning. The insights gained from this study pave the way for further research and development in optimizing liver surgery strategies for improved patient outcomes.

\section{Conclusion}

Liver resection is a complex and challenging surgical procedure that requires precise planning and execution. Computer-aided pre-operative planning systems have been developed to assist surgeons in preparing for liver surgery. Our work has contributed to the field of liver surgery planning by presenting a novel method for computing and visualizing virtual resections, tailored to atypical resections. The implementation is based on 3D Slicer and it is available through the 3D Slicer extension manager. With the use of this software, we enable surgeons to make informed decisions based on individual patient needs and different surgical strategies. In this work, the described method has been used to evaluate parenchymal-sparing resections and anatomical resections.
Our findings showed that PSR using our proposed method resulted in better preservation of parenchyma than AR resection using the centerline extraction-based method. Summarized, our study contributes to the development of effective surgical planning methods for liver resections. Future work will expand upon this study by incorporating a new dataset that includes intra-operative information, involving a larger participant cohort and examining how our planning tool could be integrated into surgical procedures during operations.

\section*{Acknowledgments}
The authors would like to express their sincere gratitude to Prof. Michael S. Floater, University of Oslo, Norway for his extensive guidance on spline theory, which fostered a vibrant and stimulating scientific exchange bridging mathematical theory and clinical applications.\\
This work was supported by the Research Council of Norway through the ALive project (311393).

\section*{Declaration of competing interest}
The authors declare that they have no known competing financial interests or personal relationships that could have appeared to influence the work reported in this paper.

\section*{Summary Points}
\paragraph{What Was Already Known on the Topic}

\begin{itemize}
	\item Surgical resection is recognized as the primary curative treatment for primary and secondary liver cancer. The planning for these resections relies heavily on computer-assisted systems that utilize virtual resection based on different geometric modeling techniques.
           
\item The advent of computer-aided pre-operative planning systems has significantly contributed to assisting surgeons by providing 3D models generated from CT/MRI data, facilitating the evaluation of potential surgical outcomes through the modeling of various surgical scenarios.

	\item The literature differentiates between parenchyma-sparing resection (PSR) and anatomical resection (AR), with PSR being preferred for its potential to preserve liver volume and reduce postoperative complications under specific circumstances. However, the pre-operative outcomes of these two strategies have not been previously compared using computer-aided pre-operative planning systems.
	\end{itemize}

\paragraph{What This Study Added to Our Knowledge}

\begin{itemize}
	\item This research introduced a novel approach for defining virtual resections through the application of contours placed/extracted from the liver surface and spline surface approximation. This method aligns closely with the surgical cutting path used in real surgeries, providing curved resection plans that cater to the diverse shapes resulting from PSR plans.
	\item Through the application of the newly developed method, the study embarked on a comparative analysis between PSR and AR plans. It highlighted significant differences in resected volumes and remnant liver volumetry, underscoring PSR's lower resected volumes and higher remnant liver volumes compared to AR. This emphasizes the efficacy and potential advantages of PSR in optimizing liver resection strategies.
	\item The findings and quantitative assessments of this study contribute towards refining the surgical planning process, particularly by highlighting the benefits of PSR in preserving healthy liver tissue. It emphasizes the importance of detailed, patient-specific virtual resection planning in enhancing the surgical decision-making process for liver surgeries.
\end{itemize}

\bibliographystyle{elsarticle-num} 
\bibliography{cas-refs}

\end{document}